\documentclass[paper]{ieice}
\usepackage{graphicx,xcolor}
\usepackage{amsmath}
\usepackage{comment}
\usepackage{cite}
\usepackage{bm}
\usepackage{subfigure}

\field{A}
\title{A hue-preserving tone mapping scheme\\based on constant-hue plane without gamut problem}
\authorlist{%
 \authorentry{Yuma KINOSHITA}{s}{A}\MembershipNumber{1614118}
 \authorentry{Kouki SEO}{n}{A}\MembershipNumber{}
 \authorentry{Artit VISAVAKITCHAROEN}{n}{A}\MembershipNumber{}
 \authorentry[kiya@tmu.ac.jp]{Hitoshi KIYA}{f}{A}\MembershipNumber{7904021}
}
\affiliate[A]{The authors are with Department of Information and Commu
nication Systems, Tokyo Metropolitan University, Hino-shi, 191-0065 Japan.
}

\received{2019}{1}{1}
\revised{2019}{1}{1}

\begin{document}
\setlength{\textfloatsep}{1pt}
\setlength{\floatsep}{1pt}
\setlength{\abovedisplayskip}{1pt}
\setlength{\belowdisplayskip}{1pt}
\setlength{\abovecaptionskip}{1pt}
\setlength{\belowcaptionskip}{1pt}
\setlength{\tabcolsep}{1pt}
\setlength{\dblfloatsep}{1pt}
\setlength{\dbltextfloatsep}{1pt}

\maketitle
\begin{summary}
We propose a novel hue-preserving tone mapping scheme.
Various tone mapping operations have been studied so far, but there are very few works on color distortion caused in image tone mapping.
First, LDR images produced from HDR ones by using conventional tone mapping operators (TMOs) are pointed out to have some distortion in hue values due to clipping and rounding quantization processing.
Next,we propose a novel method which allows LDR images to have the same maximally saturated color values as those of HDR ones.
Generated LDR images by the proposed method have smaller hue degradation than LDR ones generated by conventional TMOs.
Moreover, the proposed method is applicable to any TMOs.
In an experiment, the proposed method is demonstrated not only to produce images with small hue degradation but also to maintain well-mapped luminance, in terms of  three objective metrics: TMQI, hue value in CIEDE2000, and the maximally saturated color on the constant-hue plane in the RGB color space.  
\end{summary}

\section{Introduction}
\label{sec:intro}

The interest of high dynamic range (HDR) imaging has recently been increasing
in various area: photography, medical imaging, computer graphics,
on vehicle cameras, astronautics.
HDR images have the information of the wide dynamic range of real scenes.
However, commonly used display devices having low dynamic range (LDR)
cannot express the information that HDR images have.
In contrast, display devices which can directly represent HDR images
are not popular yet.
Therefore, the importance of tone mapping (TM) operations has been growing
in order to display HDR images on conventional LDR devices.

Tone mapping (TM) operation generates LDR images from HDR ones to display HDR ones on conventional LDR devices.
Various research works on TM have so far been reported
\cite{Reinhard,Drago,Fattal,Shan,Gu,Durand,Siku,tone1,tone2,tone3,tone4,tone5,tone6}.
Most conventional studies have focused on compressing the luminance range of HDR images.
In these methods, the range of the luminance of an HDR image is compressed by using a TM operator (TMO),
and then an LDR image is produced by combining the compressed luminance
and the color information of the original HDR image.
However, TM operations only focusing on the luminance cause image colors to be
distorted as pointed out in \cite{Mantiuk}.
Mantiuk et al. \cite{Mantiuk} showed that
the color distortion in tone-mapped LDR images occur due a tone curve,
and they proposed a color correction formula on the basis of
experimental results illustrating a relationship
between the contrast-compression ratio and the saturation of the image.
However, they only focused on tone curves as the cause of the color distortion,
and other causes have never been considered.

In this paper, we first point out that
color distortion in tone-mapped images is primarily caused by clipping and quantizing pixel values of tone-mapped LDR images.
Then, we propose a novel hue-preserving tone mapping scheme based on constant-hue plane in the RGB color space.
The proposed method consists of two steps:
tone mapping with an conventional method and compensating hue distorted by the tone mapping.
The proposed compensation method utilizes the hue information
based on the maximally saturated colors \cite{Yamaguchi,Artit},
for suppressing color distortions due to the tone mapping.
The compensation is done by replacing the hue information of a tone-mapped LDR image
with that of the original HDR image.
Hue-preserving image-processing methods have already been developed
in various research areas such as image enhancement and noise reduction
\cite{Yamaguchi,ie1,ie2,ie3,ie4,ie5,ie6}.
These methods perform some operations under the conditions that the hue is fixed.
In contrast, there are very few hue-preserving TM operations \cite{Mantiuk},
and the proposed hue compensation method is independently performed from TM.
This approach enables us not only to reduce color distortion, but also to maintain advantage of conventional TMOs.

To evaluate the effectiveness of the proposed method,
we performed a number of simulations.
In the simulations, the proposed method was compared with conventional TM operations
in terms of the hue difference used in CIEDE 2000 \cite{CIEDE2000}
and the maximally saturated color difference.
Experimental results showed that the hue difference between tone-mapped LDR images
and the original HDR images can be reduced by applying the proposed method.
Moreover, results of objective quality evaluation
with the tone mapped image quality index (TMQI) \cite{TMQI} illustrated that
the proposed method can maintain the performance of conventional TMOs.

\section{PREPARATION}
\label{sec:pre}

A general TM operation and the constant-hue plane in the RGB color space are summarized, here.

\subsection{Tone mapping}
\label{ssec:tm}

A general TM operation is summarized, here. A TM operation consists of the following four steps.
\begin{description}
\renewcommand{\labelenumi}{\alph{enumi}).}
  \item[(a)] The world luminance $L_w(p)$ of an HDR image $I_H$ is
calculated from RGB pixel values of the HDR image
as,
\end{description}
\begin{equation}
\label{Lw}
L_w(p)=0.27R(p)+0.67G(p)+0.06B(p)
\end{equation}
where $R(p),G(p)$ and $B(p)$ are RGB pixel values of the HDR image with a pixel $p$, respectively.
\begin{description}
\renewcommand{\labelenumi}{\alph{enumi}).}
  \item[(b)] The display luminance $L_d(p)$ is calculated by using a
TMO.
\end{description}
\begin{description}
\renewcommand{\labelenumi}{\alph{enumi}).}
  \item[(c)] The floating-point pixel values $C_f(p)$ of the LDR image
is calculated as follows:
\end{description}
\begin{equation}
\label{Cf}
C_f(p)=\frac{L_d(p)}{L_w(p)}C(p)
\end{equation}
where $C(p) \in \{R(p),G(p),B(p)\}$ is the floating-point
RGB value of the input HDR image $I_H$ , and $C_f(p) \in \{R_f(p),G_f(p),B_f(p)\}$. Besides, the gamma correction is performed for $C_f(p)$ as needed.
\begin{description}
\renewcommand{\labelenumi}{\alph{enumi}).}
  \item[(d)] The 8-bit color RGB values $C_i(p)$ of the LDR image $I_L$
is derived from
\end{description}
\begin{equation}
\label{round}
C_i(p)=\mathrm{round}(C_f \cdot 255)
\end{equation}
where $\mathrm{round}(A)$ rounds $A$ to its nearest integer value,
and $C_i(p) \in \{R_i(p),G_i(p),B_i(p)\}$.
Further, the value of $C_i(p)$ is redefined as follows:
\begin{equation}
\label{clip}
C_i(p)= \left \{
\begin{array}{l}
0 \quad (C_i(p)<0) \\
C_i(p) \quad (0 \leq C_i(p) \leq 255) \\
255 \quad (255<C_i(p))
\end{array}
\right.
\end{equation} 
where if $C_i(p)$ takes a value over 255, it is clipped at 255.
Also, if $C_i(p)$ takes a negative value, it is clipped at 0.
The rounding quantization in Eq.(\ref{round}) and clipping in Eq.(\ref{clip}) generate some error.

In the case of "Photographic Tone Reproduction" \cite{Reinhard} which is a typical TMO, the display luminance $L_d(p)$ in step(b) is calculated in accordance with the following procedure.
The geometric mean $\overline{L}_w$ of the world luminance $L_w(p)$ is calculated as follows:
\begin{equation}
\overline{L}_w=\exp(\frac{1}{N}\sum_{p=1}^{N} \log L_w(p))
\end{equation}
where $N$ is the total number of pixels in the input HDR image $I_H$.
The scaled luminance $L(p)$ is calculated as
\begin{equation}
L(p)=\frac{\alpha}{\overline{L}_w}L_w(p)
\end{equation}
where $\alpha \in[0,1]$ is the parameter called “key value”,
which indicates subjectively if the scene is light, normal, or dark.
The display luminance $L_d (p)$ is calculated by using the
TMO as follows:
\begin{equation}
L_d(p)=\frac{L(p)}{1+L(p)}
\end{equation}

\subsection{Constant hue plane in the RGB color space}
\label{ssec:hueplane}

\begin{figure}[t]
  \centering
\includegraphics[height=40mm]{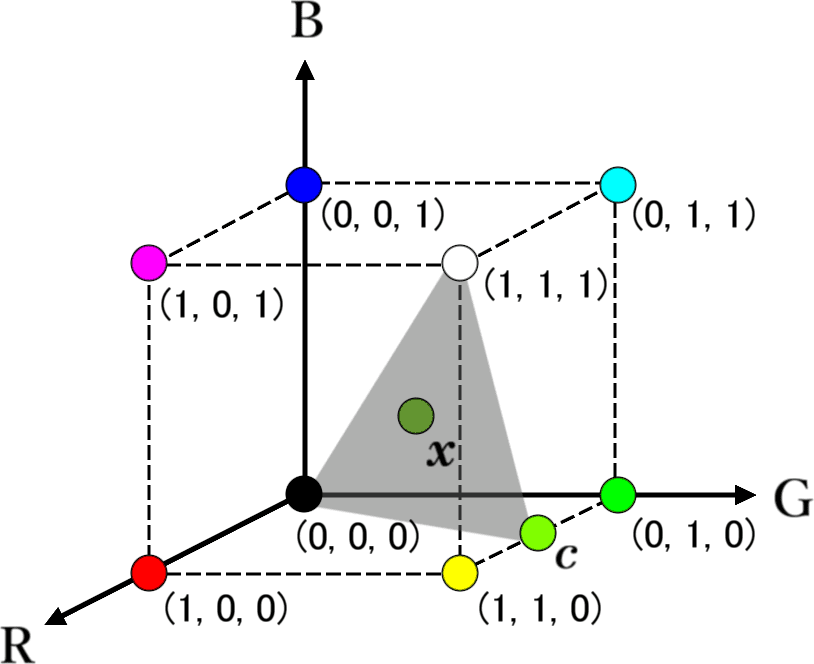}
\caption{A conceptual diagram of the RGB color space}
\end{figure} 

We focus on the constant hue plane in the RGB color space \cite{Yamaguchi} to discuss color distortion. 
An input image is a 24-bit full color image and each pixel of the image is represented as $\bm{x}\in[0,1]^3$. 
$x_r, x_g$ and $x_b$ are the R, G, and B components of the pixel $\bm{x}$, respectively, as shown in Fig.1.
In the RGB color space, a set of pixels which has the same hue forms a plane, called constant hue plane.
The shape of the constant hue plane is the triangle whose vertices correspond to white, black and the maximally saturated color,
where $\bm{w}=(1,1,1)$, $\bm{k}=(0,0,0)$ and $\bm{c}$ are white, black and the maximally saturated color with the same hue as $\bm{x}$, respectively. 
The maximally saturated color $\bm{c}=(c_r,c_g,c_b)$ is calculated by
\[
c_r = \frac{x_r-\min{(\bm{x})}}{\max{(\bm{x})}-\min{(\bm{x})}},
\]
\begin{equation}
\label{c}
c_g = \frac{x_g-\min{(\bm{x})}}{\max{(\bm{x})}-\min{(\bm{x})}},
\end{equation}
\[
c_b = \frac{x_b-\min{(\bm{x})}}{\max{(\bm{x})}-\min{(\bm{x})}}
\]
where $\max{(\cdot)}$ and $\min{(\cdot)}$ are functions that return the maximum and minimum elements of the pixel $\bm{x}$, respectively.
Therefore, from Eq.(\ref{c}), the elements of $\bm{c}$ corresponding to the maximum and minimum elements of the pixel $\bm{x}$ become 1 and 0, respectively.

On the constant hue plane, a pixel $\bm{x}$ can be represented as a linear combination as
\begin{equation}
\label{x}
\bm{x}=a_w \bm{w}+a_k \bm{k}+a_c \bm{c}
\end{equation}
where 
\[
a_w = \min(\bm{x}),
\]
\begin{equation}
\label{a}
a_c = \max(\bm{x})-\min(\bm{x}),
\end{equation}
\[
a_k = 1-\max(\bm{x}).
\]
Since $\bm{w},\bm{k},\bm{c}$ and $\bm{x}$ exist on the plane and $\bm{x}$ is an interior point of $\bm{w},\bm{k}$ and $\bm{c}$, the following equations hold.
\begin{equation}
\label{a1}
a_w +a_k+a_c=1,
\end{equation}
\begin{equation}
\label{a2}
0 \leq a_w,a_k,a_c \leq 1.
\end{equation}

\section{PROPOSED HUE COMPENSATION}
\label{sec:proposal}

We propose a method for compensating hue values of tone-mapped images.

\subsection{Hue distortion}
\label{ssec:causes}

Hue distortion occurs due to the influence of the following three operations.
\begin{description}
\renewcommand{\labelenumi}{\alph{enumi}).}
\setlength{\leftskip}{0.3cm}
  \item[a)] Tone curve in step (b)
  \item[b)] Rounding quantization in Eq.(\ref{round})
  \item[c)] Clipping in Eq.(\ref{clip})
\end{description}
Mantiuk et al. pointed out that a) tone curve generates color distortion on the CIECAM02 color appearance model, and proposed a correction formula \cite{Mantiuk}. 
However, they have never pointed out the influence of the operations b) and c).

The framework of the proposed method is shown in Fig.2.
As illustrated in Fig.2, the proposed method can be applied to any TMO and moreover the influence of the above three operations can be considered.

\begin{figure}[t]
  \centering
\includegraphics[width=80mm]{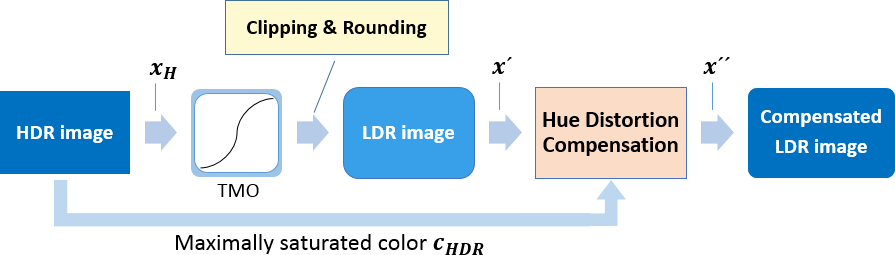}
\caption{The framework of the proposed method}
\end{figure}

Figure 3 shows an example of the difference of the maximally saturated colors between an original HDR image and the tone-mapped LDR image including the influence of clipping and rounding quantization processing. 
Figure 4 shows an example of the maximally saturated color of an original HDR image and the tone-mapped LDR image.
In these figures, the influence of rounding quantization and clipping is demonstrated to cause color distortion.

\begin{figure*}[h]
\centering
  \subfigure[]{%
       \includegraphics[clip, height=0.46\columnwidth]{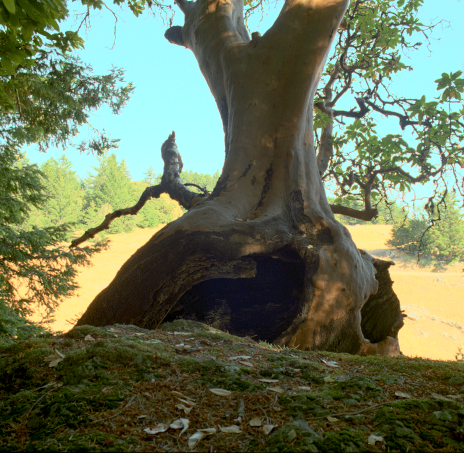}}%
\hspace{0.01mm}
　　\subfigure[]{%
       \includegraphics[clip, height=0.46\columnwidth]{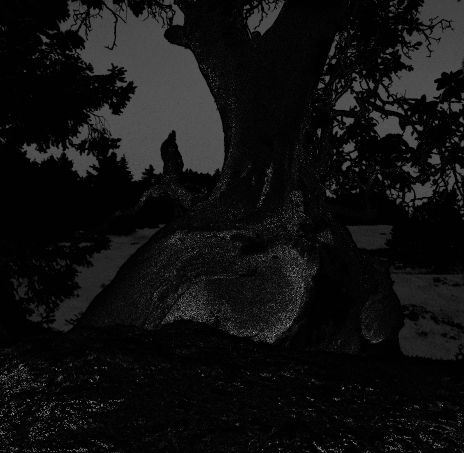}}%
\hspace{0.01mm}
  \subfigure[]{%
       \includegraphics[clip, height=0.46\columnwidth]{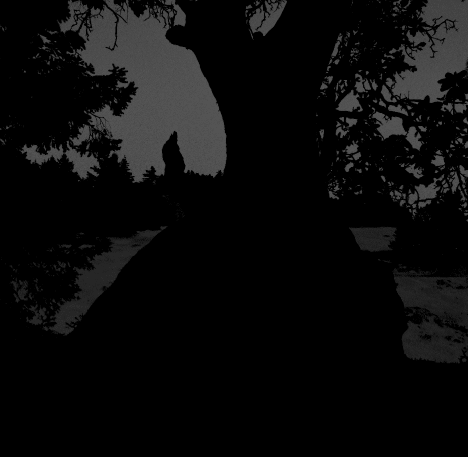}}%
\hspace{0.01mm}
  \subfigure[]{%
       \includegraphics[clip, height=0.458\columnwidth]{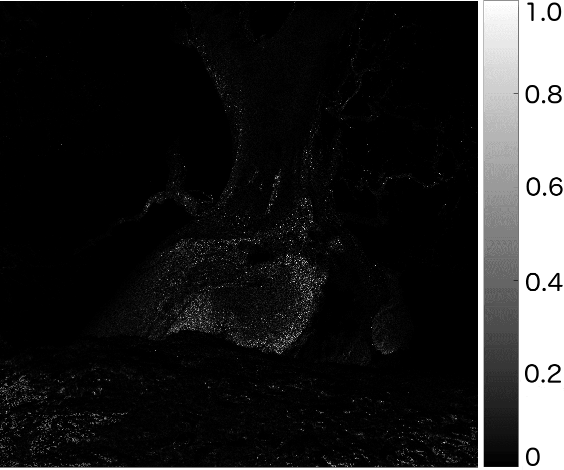}}%
  \caption{Difference of maximally saturated colors in "Tree" where the reference is the maximally saturated color of the original HDR image. (a) Tone-mapped image, (b) Difference (performed clipping \& rounding), (c) Difference (performed only clipping), (d) Difference (performed only rounding)}
\end{figure*}

\begin{figure*}[h]
\centering
  \subfigure[]{%
       \includegraphics[clip, height=0.50\columnwidth]{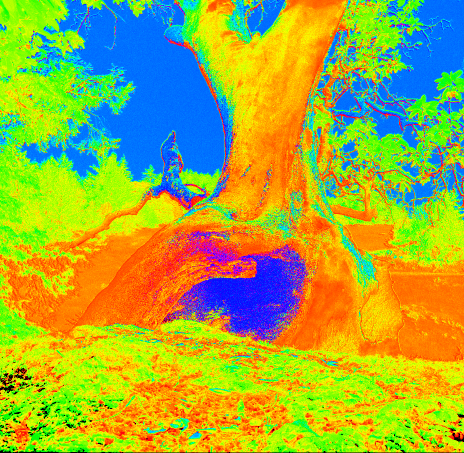}}%
  \hspace{10mm}
　　\subfigure[]{%
       \includegraphics[clip, height=0.50\columnwidth]{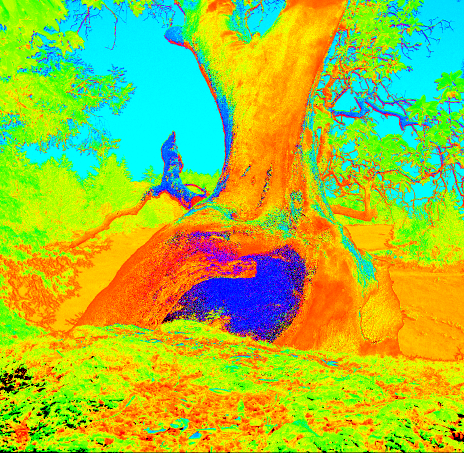}}%
  \caption{Maximally saturated color in "Tree". (a) Original HDR image, (b) Tone-mapped LDR image with clipping and rounding quantization}
\end{figure*}

\subsection{HDR images on the constant hue plane}
\label{ssec:causes}

We consider giving HDR images the same pixel representation on the constant hue plane as LDR ones described in 3.1.

Each pixel of an HDR image is represented as $\bm{x}_H=(x_r, x_g, x_b)$ where $x_r, x_g$ and $x_b$ are generally real numbers.
As well as LDR images, we define white, black and maximally saturated color as $\bm{w}=(1,1,1)$, $\bm{k}=(0,0,0)$ and $\bm{c}_{H}$.
Maximally saturated color $\bm{c}_{H}$ of the HDR image is calculated by replacing $\bm{x}$ with $\bm{x}_{H}$ in Eq.(\ref{c}), and then the pixel value $\bm{x}_H$ is represented as a linear combination as
\begin{equation}
  \bm{x}_H=a_w \bm{w}+a_k \bm{k}+a_c \bm{c}_{H}
\end{equation}
where $a_w, a_k$ and $a_c$ are coefficients which are calculated in accordance with Eq.(\ref{a}).
For HDR images, $\bm{w},\bm{k},\bm{c}_H$ and $\bm{x}_H$ exist on the same plane in the RGB color space, but $\bm{x}_H$ is not always an interior point of $\bm{w},\bm{k}$ and $\bm{c}_H$. That is, $a_w, a_k$ and $a_c$ do not meet Eq.(\ref{a2}), although they satisfy Eq.(\ref{a1}).

The proposed method aims to suppress some hue distortion included in LDR images mapped from HDR ones.
Now, let $\bm{x}'$ be a pixel value of a tone-mapped LDR image that includes the influence of rounding quantization and clipping processing.
In accordance with Eq.(\ref{x}), the pixel value $\bm{x}'$ is expressed as follows
\begin{equation}
\label{x'}
  \bm{x}'= a'_w \bm{w}+ a'_k \bm {k}+ a'_c \bm {c}'
\end{equation}
where $ \bm{c}'$ is a maximally saturated color calculated from the pixel
value $\bm{x}'$ by Eq.(\ref{c}).
$ a'_w, a'_k$ and $a'_c $ are coefficients calculated by Eq.(\ref{a}).
Here, in general, $ \bm{c}'= \bm{c}_H $ is not met due to the above
reasons. 
Therefore, to correct the hue of $\bm{x}'$, we replace $\bm{x}'$ with $\bm{x}''$ as,
\begin{equation}
\label{x''}
  \bm{x}'' = a'_w \bm {w}+ a'_k \bm{k}+ a'_c \bm{c}_{H}.
\end{equation}
Note that $\bm{x}_H$ and $\bm{x}''$ are on the same constant hue plane.

\subsection{Proposed procedure}

The procedure of the proposed tone mapping scheme is summarized as follows.
\begin {description}
  \item [(1)] Generate an LDR image from an HDR image by using a conventional TMO.
  \item [(2)] Calculate the coefficients $a'_w, a'_k, a'_c$ of the generated LDR image for each pixel value $\bm{x}'$ in accordance with Eq.(\ref{a}).
  \item [(3)] Calculate the maximally saturated color $ \bm{c}_{H} $ of the original HDR image by using Eq.(\ref{c}).
  \item[(4)] Calculate the pixel value $\bm{x}''=(x''_r, x''_g, x''_b)$ of the compensated LDR image in accordance with Eq.(\ref{x''}).
\end{description}
The R, G and B components of $ \bm{c}_{H} $ are in the range [0, 1] from Eq.(\ref{c}). Therefore, the R, G and B components of $ \bm {x}'' $ are also in the range [0, 1] even after compensation.
This allows us to prevent clipping error.

The proposed scheme is applicable to any conventional TMO.
In addition, it can consider all influences caused in three operations described in 3.1.

\section {Simulation}
In an experiment, the proposed scheme was compared with conventional TMOs without any compensation and Mantiuk's hue correction formula.

\subsection{Simulation condition}

We used 10 HDR images selected from an HDR image database \cite{HDR} for the evaluation.
The following is the procedure of the evaluation.
\begin {description}
\renewcommand {\labelenumi} {\alph{enumi}).}
  \item [(1)] Generate LDR images $I'_L$ from prepared HDR images $I_H$ by using four TMOs: Reinhard's Global Operator \cite{Reinhard}, Reinhard's Local Operator \cite{Reinhard}, Drago's TMO \cite{Drago}, Fattal's TMO \cite{Fattal}, Shan's TMO\cite{Shan} and Gu's TMO\cite{Gu}.
  \item [(2)] Apply the proposed method to $I'_L$, and obtain compensated LDR images $I''_L$.
  \item [(3)] Compare $I''_L$ with $I'_L$ and images corrected by Mantiuk's method.
\end{description}

\subsection {Simulation results}
\renewcommand \thesubsubsection {\Alph{subsubsection}}

\begin{table*}[t]
\centering
\caption{Simulation result ($\Delta \bm{c}$)}
  {
\scriptsize
\begin{tabular}{l|cc|cc|cc|cc|cc|cc}
\hline
\multicolumn{1}{c|}{}       & \multicolumn{2}{c|}{Reinhard(global)} & \multicolumn{2}{c|}{Reinhard(local)} & \multicolumn{2}{c|}{Drago} & \multicolumn{2}{c|}{Fattal} & \multicolumn{2}{c|}{Shan} & \multicolumn{2}{c}{Gu}   \\ \cline{2-13} 
\multicolumn{1}{c|}{Images} & Conv.         & Prop.                 & Conv.        & Prop.                 & Conv.   & Prop.            & Conv.    & Prop.            & Conv.   & Prop.           & Conv.  & Prop.           \\ \hline
Apartment\_float            & 0.0766        & \textbf{0.0672}       & 0.1405       & \textbf{0.1318}       & 0.0198  & \textbf{0.0133}  & 0.0136   & \textbf{0.0099}  & 0.0472  & \textbf{0.0400} & 0.0132 & \textbf{0.0089} \\
dani\_belgium               & 0.0130        & \textbf{0.0104}       & 0.0388       & \textbf{0.0315}       & 0.0150  & \textbf{0.0036}  & 0.0103   & \textbf{0.0036}  & 0.0274  & \textbf{0.0217} & 0.0131 & \textbf{0.0053} \\
Desk                        & 0.0227        & \textbf{0.0136}       & 0.0655       & \textbf{0.0450}       & 0.0276  & \textbf{0.0028}  & 0.0166   & \textbf{0.0030}  & 0.0439  & \textbf{0.0320} & 0.0238 & \textbf{0.0057} \\
Display1000\_float          & 0.0222        & \textbf{0.0184}       & 0.0830       & \textbf{0.0737}       & 0.0137  & \textbf{0.0050}  & 0.0127   & \textbf{0.0076}  & 0.0269  & \textbf{0.0208} & 0.0170 & \textbf{0.0107} \\
memorial                    & 0.0605        & \textbf{0.0429}       & 0.1116       & \textbf{0.0973}       & 0.0345  & \textbf{0.0054}  & 0.0308   & \textbf{0.0016}  & 0.0321  & \textbf{0.0028} & 0.0362 & \textbf{0.0032} \\
MtTamWest                   & 0.0053        & \textbf{0.0024}       & 0.0449       & \textbf{0.0121}       & 0.0468  & \textbf{0.0009}  & 0.0175   & \textbf{0.0081}  & 0.0406  & \textbf{0.0300} & 0.0205 & \textbf{0.0065} \\
rend06                      & 0.0159        & \textbf{0.0037}       & 0.0463       & \textbf{0.0322}       & 0.0204  & \textbf{0.0016}  & 0.0396   & \textbf{0.0016}  & 0.0411  & \textbf{0.0025} & 0.0518 & \textbf{0.0030} \\
rosette                     & 0.0056        & \textbf{0.0035}       & 0.0558       & \textbf{0.0062}       & 0.0640  & \textbf{0.0008}  & 0.0106   & \textbf{0.0023}  & 0.0127  & \textbf{0.0044} & 0.0127 & \textbf{0.0025} \\
StillLife                   & 0.0221        & \textbf{0.0029}       & 0.0507       & \textbf{0.0190}       & 0.0355  & \textbf{0.0007}  & 0.0247   & \textbf{0.0009}  & 0.0258  & \textbf{0.0020} & 0.0316 & \textbf{0.0022} \\
Tree                        & 0.0434        & \textbf{0.0152}       & 0.0844       & \textbf{0.0622}       & 0.0435  & \textbf{0.0029}  & 0.0213   & \textbf{0.0045}  & 0.0216  & \textbf{0.0066} & 0.0223 & \textbf{0.0022} \\ \hline
\end{tabular}
 }
\end{table*}

\begin{table*}[t]
\centering
\caption{Simulation result ($\Delta \mathrm{H}$)}
  {
\scriptsize
\begin{tabular}{l|cc|cc|cc|cc|cc|cc}
\hline
\multicolumn{1}{c|}{}       & \multicolumn{2}{c|}{Reinhard(global)} & \multicolumn{2}{c|}{Reinhard(local)} & \multicolumn{2}{c|}{Drago} & \multicolumn{2}{c|}{Fattal}       & \multicolumn{2}{c|}{Shan}         & \multicolumn{2}{c}{Gu}    \\ \cline{2-13} 
\multicolumn{1}{c|}{Images} & Conv.         & Prop.                 & Conv.         & Prop.                & Conv.    & Prop.           & Conv.           & Prop.           & Conv.           & Prop.           & Conv.   & Prop.           \\ \hline
Apartment\_float            & 0.8193        & \textbf{0.2674}       & 0.4721        & \textbf{0.3147}      & 0.4177   & \textbf{0.1720} & 0.2542          & \textbf{0.1514} & 0.4268          & \textbf{0.2959} & 0.2780  & \textbf{0.1281} \\
dani\_belgium               & 0.5589        & \textbf{0.3940}       & 0.9748        & \textbf{0.7122}      & 0.9940   & \textbf{0.4174} & 0.6488          & \textbf{0.3175} & 0.9644          & \textbf{0.6987} & 0.6961  & \textbf{0.3003} \\
Desk                        & 2.6314        & \textbf{0.6395}       & 3.3420        & \textbf{1.3861}      & 3.9240   & \textbf{0.9566} & 2.0510          & \textbf{0.7582} & 2.1173          & \textbf{1.2641} & 2.9466  & \textbf{0.7737} \\
Display1000\_float          & 0.4630        & \textbf{0.2664}       & 0.9950        & \textbf{0.7335}      & 0.6737   & \textbf{0.2687} & 0.4691          & \textbf{0.2634} & 0.6964          & \textbf{0.4213} & 0.5600  & \textbf{0.2615} \\
memorial                    & 3.2015        & \textbf{0.4858}       & 1.8574        & \textbf{0.7741}      & 3.5305   & \textbf{0.2863} & 3.8368          & \textbf{1.5534} & 4.5643          & \textbf{1.3001} & 4.8098  & \textbf{1.6655} \\
MtTamWest                   & 1.0522        & \textbf{0.5513}       & 4.0217        & \textbf{1.8243}      & 6.9016   & \textbf{1.9800} & 1.1497          & \textbf{0.2810} & 1.4351          & \textbf{0.5257} & 1.8670  & \textbf{0.2307} \\
rend06                      & 1.9946        & \textbf{0.5657}       & 2.0995        & \textbf{1.4428}      & 1.7591   & \textbf{1.1248} & 8.4673          & \textbf{3.7136} & 9.3332          & \textbf{2.8647} & 13.5360 & \textbf{3.5505} \\
rosette                     & 0.7701        & \textbf{0.4208}       & 10.3067       & \textbf{4.9597}      & 18.1123  & \textbf{2.9129} & \textbf{0.8894} & 0.9286          & \textbf{0.8983} & 0.9128          & 1.1258  & \textbf{0.9651} \\
StillLife                   & 7.9052        & \textbf{2.5178}       & 6.8135        & \textbf{2.8522}      & 11.3119  & \textbf{4.9404} & 6.8368          & \textbf{3.8103} & 6.6331          & \textbf{3.5000} & 9.9990  & \textbf{4.8098} \\
Tree                        & 6.4565        & \textbf{0.5583}       & 3.2179        & \textbf{1.0439}      & 5.6125   & \textbf{0.9509} & 2.1995          & \textbf{0.7863} & 1.9026          & \textbf{0.7583} & 2.7228  & \textbf{0.8890} \\ \hline
\end{tabular}
 }
\end{table*}

\begin{table*}[t]
\centering
\caption{Simulation result (TMQI)}
  {
\scriptsize
\begin{tabular}{l|cc|cc|cc|cc|cc|cc}
\hline
\multicolumn{1}{c|}{}       & \multicolumn{2}{c|}{Reinhard(global)} & \multicolumn{2}{c|}{Reinhard(local)} & \multicolumn{2}{c|}{Drago}        & \multicolumn{2}{c|}{Fattal}       & \multicolumn{2}{c|}{Shan}         & \multicolumn{2}{c}{Gu}            \\ \cline{2-13} 
\multicolumn{1}{c|}{Images} & Conv.             & Prop.             & Conv.             & Prop.            & Conv.           & Prop.           & Conv.           & Prop.           & Conv.           & Prop.           & Conv.           & Prop.           \\ \hline
Apartment\_float            & 0.8024            & \textbf{0.8029}   & 0.8235            & \textbf{0.8241}  & 0.7935          & \textbf{0.7941} & 0.7887          & \textbf{0.7887} & 0.7309          & \textbf{0.7310} & 0.8892          & \textbf{0.8894} \\
dani\_belgium               & \textbf{0.9219}   & 0.9214            & \textbf{0.9011}   & 0.9009           & 0.9072          & \textbf{0.9072} & 0.8542          & \textbf{0.8542} & 0.7368          & \textbf{0.7369} & 0.9489          & \textbf{0.9508} \\
Desk                        & \textbf{0.9609}   & 0.9535            & \textbf{0.9495}   & 0.9407           & \textbf{0.9604} & 0.9443          & \textbf{0.8921} & 0.8815          & \textbf{0.7521} & 0.7510          & 0.8272          & \textbf{0.8335} \\
Display1000\_float          & \textbf{0.9551}   & 0.9537            & \textbf{0.8609}   & 0.8596           & \textbf{0.9370} & 0.9363          & \textbf{0.7882} & 0.7873          & \textbf{0.7316} & 0.7311          & \textbf{0.8795} & 0.8782          \\
memorial                    & 0.8271            & \textbf{0.8475}   & 0.8563            & \textbf{0.8601}  & 0.8325          & \textbf{0.8590} & \textbf{0.7978} & 0.7909          & \textbf{0.7894} & 0.7828          & 0.8553          & \textbf{0.8701} \\
MtTamWest                   & \textbf{0.9673}   & 0.9643            & \textbf{0.8936}   & 0.8797           & \textbf{0.9518} & 0.9185          & \textbf{0.8052} & 0.8039          & \textbf{0.7798} & 0.7795          & 0.9267          & \textbf{0.9347} \\
rend06                      & \textbf{0.9491}   & 0.9441            & \textbf{0.8149}   & 0.8108           & \textbf{0.9385} & 0.9277          & \textbf{0.7873} & 0.7758          & \textbf{0.7840} & 0.7644          & \textbf{0.9218} & 0.8241          \\
rosette                     & \textbf{0.8265}   & 0.8231            & \textbf{0.8469}   & 0.8304           & \textbf{0.8136} & 0.7566          & \textbf{0.7968} & 0.7936          & \textbf{0.7632} & 0.7605          & \textbf{0.8741} & 0.8704          \\
StillLife                   & \textbf{0.8825}   & 0.8436            & \textbf{0.8211}   & 0.7996           & \textbf{0.8611} & 0.8117          & 0.7272          & \textbf{0.7284} & \textbf{0.7423} & 0.7384          & \textbf{0.8230} & 0.8011          \\
Tree                        & \textbf{0.9609}   & 0.9430            & \textbf{0.9212}   & 0.9117           & \textbf{0.9641} & 0.9473          & \textbf{0.8499} & 0.8407          & \textbf{0.7748} & 0.7701          & 0.8340          & \textbf{0.8381} \\ \hline
\end{tabular}
 }
\end{table*}

\begin{table*}[t]
\centering
\caption{Comparison with Mantiuk's method}
  {
    \scriptsize
\begin{tabular}{l|ccc|ccc}
\hline
\multicolumn{1}{c|}{}    & \multicolumn{3}{c|}{$\Delta \bm{c}$}                & \multicolumn{3}{c}{$\Delta \mathrm{H}$}               \\ \cline{2-7} 
\multicolumn{1}{c|}{Images} & Conventional  & Mantiuk \cite{Mantiuk} & Proposed             & Conventional  & Mantiuk \cite{Mantiuk} & Proposed             \\ \hline
dani\_belgium         & 0.0505 & 0.0548      & \textbf{0.0467}  & 1.8327 & 1.8655      & \textbf{1.6308}  \\
memorial              & 0.0490 & 0.0548      & \textbf{0.0184}  & 5.3609 & 5.5372      & \textbf{3.0708}  \\
Tree                  & 0.1123 & 0.1151      & \textbf{0.0894}  & 6.7300 & 6.0663      & \textbf{3.3379}  \\ \hline
\end{tabular}
  }
\end{table*}

\subsubsection {Hue distortion}
We used two objective metrics to evaluate hue distortion caused in TM operations.
One is the difference in the maximally saturated colors between image $I_1$ with $M \times N$ pixels and $I_2$ with $M \times N$ pixels.
it is calculated by the following three steps.
\begin {description}
\renewcommand {\labelenumi} {\alph {enumi}).}

  \item [(1)] Calculate the maximally saturated colors $\bm{c}_1(i),\bm{c}_2(i)$ from a pixel value ​​$\bm{x}_1(i)$ in $I_1$ and $\bm{x}_2(i)$ in $I_2$, from Eq.(\ref{c}), where $\bm{x}_{j}(i)$ indicates the $i$-th pixel of images $I_j$, $j \in \{1,2\}$, and $\bm{c}_{j}(i)$ is its the maximally saturated color.
  \item [(2)] The difference $\Delta \bm{c}(i)$ between $\bm{c}_1(i)$ and $\bm{c}_2(i)$ for each pixel is given by

\[
\Delta\bm{c}(i) = \| \bm{c}_1(i) - \bm{c}_2(i) \|
\]
where $\|\cdot\|$ is euclidean norm.
  \item [(3)] Calculate the average value $\Delta \bm{c}$ of $\Delta \bm{c}(i)$ over all pixels.
\end {description}

The other is the hue differences in CIEDE2000, which was published by the CIE \cite{CIEDE2000}. The difference of hue values between two images was calculated as $\Delta \mathrm{H}(i)$ for each pixel, and as the average value of all pixels $\Delta \mathrm{H}$.

Tables 1 and 2 show the evaluation results in terms of the objective metrics, where $I_1$ is an HDR image and $I_2$ is a tone-mapped image.
"Conventional" indicates that images were generated by a conventional TMO without any hue compensation.
In the tables, the proposed method outperformed the conventional approach for all TMOs.
Therefore, the proposed method is effective for improving hue distortion included in fused images in terms of not only $\Delta \bm{c}$ but also $\Delta \mathrm{H}$.
Note that $\Delta \bm{c}$ for the proposed method does not become zero value.
This is because the rounding quantization has to be carried out again to generate integer pixel values, although clipping does not be done.
Figures 5, 6 and 7 show some examples of the simulation results.

\subsubsection {Influence on tone mapping}

Next, we evaluated the quality of tone-mapped LDR images in terms of Tone Mapped image Quality Index (TMQI) \cite{TMQI}, which is a well-known objective quality assessment algorithm for tone-mapped images by combining a multi-scale signal fidelity measure based on a modified structural similarity (SSIM) \cite{SSIM} index and a naturalness measure based on intensity statistics of natural images.
A higher TMQI score indicates higher quality.

From Table 3, the proposed scheme is shown to offer almost the same TMQI scores as those of the conventional approach, but many of the scores slightly decreased, due to  the change of luminance  caused by using the proposed method.
In contrast, the change of luminance is easily corrected by adjusting the mean luminance of an image by multiplying all luminance values by a constant.
Figure 8 shows the image of "StillLife" in which TMQI is remarkably low compared with the conventional method.
By comparing Fig.8(b) with Fig.8(c), it is confirmed that the slight change of luminance caused by applying the proposed method was easily corrected by the simple post-processing, so the TMQI score was improved, while maintaining $\Delta \bm{c}$ and $\Delta \mathrm{H}$.

\subsection {Comparison with Mantiuk's method}

Finally, our method was compared with Mantiuk's hue correction formula \cite{Mantiuk}.
We generated three LDR images by using Durand's TMO \cite{Durand} as in \cite{Mantiuk}, and both our method and Mantiuk's method were applied to the images respectively.
Table 4 shows the comparison result.
From the table, the our method is confirmed to outperform Mantiuk's method in terms of both $\Delta \bm{c}$ and $\Delta \mathrm{H}$.
This is because Mantiuk's method does not consider the influence of clipping and rounding quantization processing as described in 3.1.
In contrast, the our method can compensate hue distortion caused by three operations.

\section {Conclusion}
In this paper, we proposed a novel TM scheme that utilize maximally saturated colors calculated from original HDR image on the basis of the constant hue plane in the RGB color space.
The proposed method enables us to compensate the hue distortion of the tone-mapped LDR image caused by a conventional TMO, while maintaining TM performances that conventional TMOs have.
Simulation results showed the effectiveness of the proposed method in terms of three objective metrics: $\Delta \bm{c}$, $\Delta \mathrm{H}$ and TMQI.
Moreover, our method was compared with Mantiuk's method to show the effectiveness.

\begin{figure}[!t]
\centering
	    \subfigure[Conventional]{%
	        \includegraphics[clip, width=0.42\columnwidth]{f3a.png}}%
\hspace{2mm}
	    \subfigure[Proposed]{%
	        \includegraphics[clip, width=0.42\columnwidth]{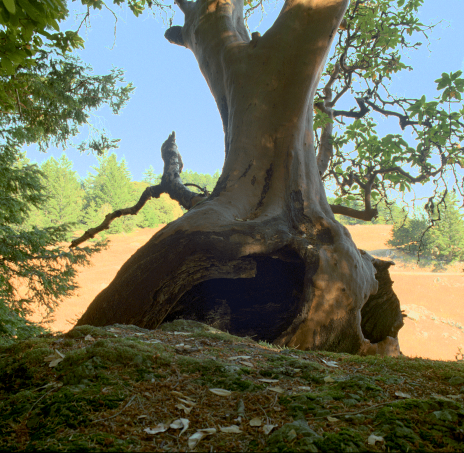}}%
	    \caption{Simulation results with Reinhard's global operator (Tree)}
\end{figure}
\begin{figure}[!t]
\centering
	    \subfigure[Conventional]{%
	        \includegraphics[clip, width=0.42\columnwidth]{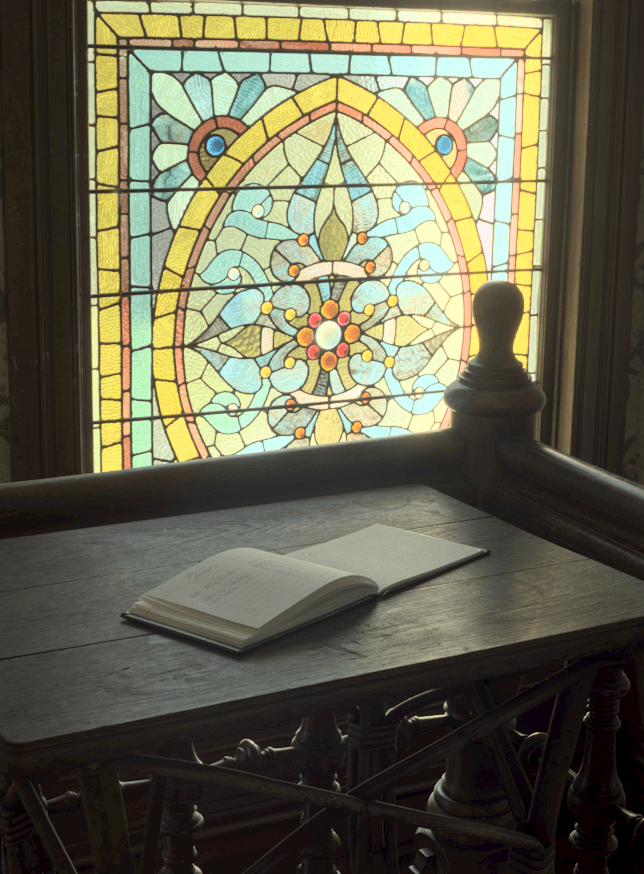}}%
\hspace{2mm}
	    \subfigure[Proposed]{%
	        \includegraphics[clip, width=0.42\columnwidth]{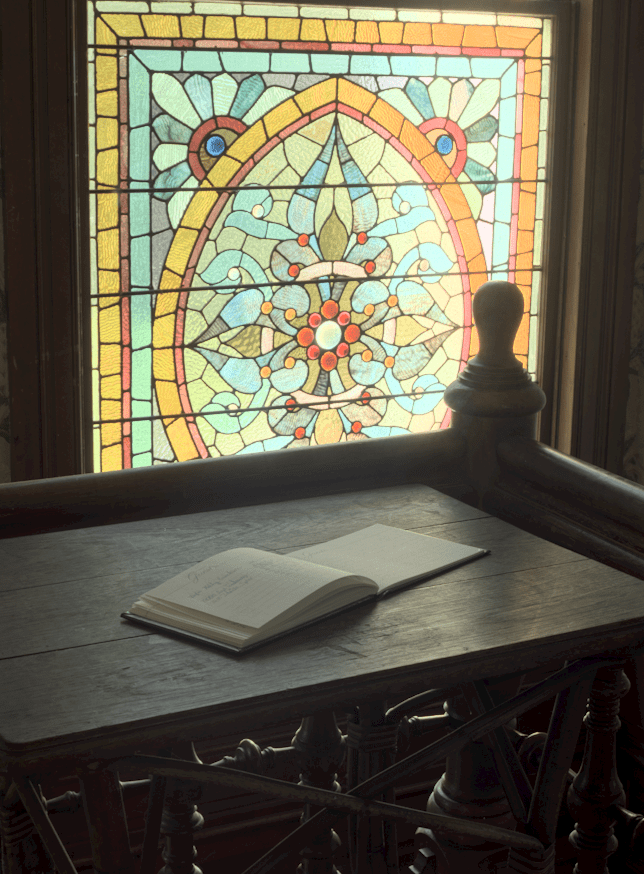}}%
	    \caption{Simulation results with Reinhard's local operator (Desk)}
\end{figure}

\begin{figure}[!t]
\centering
	    \subfigure[Conventional]{%
	        \includegraphics[clip, width=0.42\columnwidth]{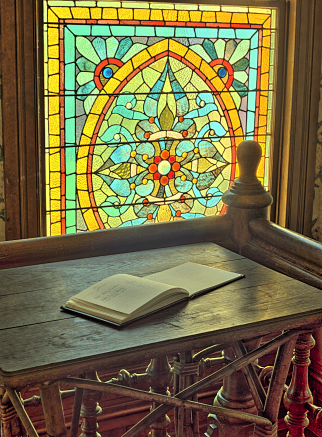}}%
\hspace{2mm}
	    \subfigure[Proposed]{%
	        \includegraphics[clip, width=0.42\columnwidth]{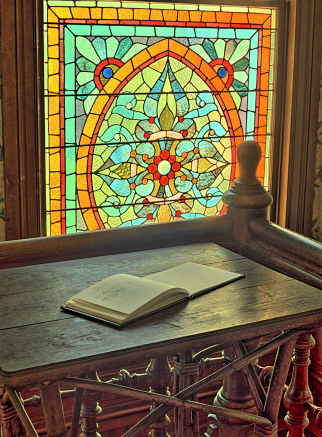}}%
	    \caption{Simulation results with Gu's TMO (Desk)}
\end{figure}

\begin{figure}[!t]
\centering
	    \subfigure[Conventional
			\protect\newline(TMQI=0.8825, $\Delta \bm{c}$=0.0221, $\Delta \mathrm{H}$=7.9052)]{%
	        \includegraphics[clip, width=0.42\columnwidth]{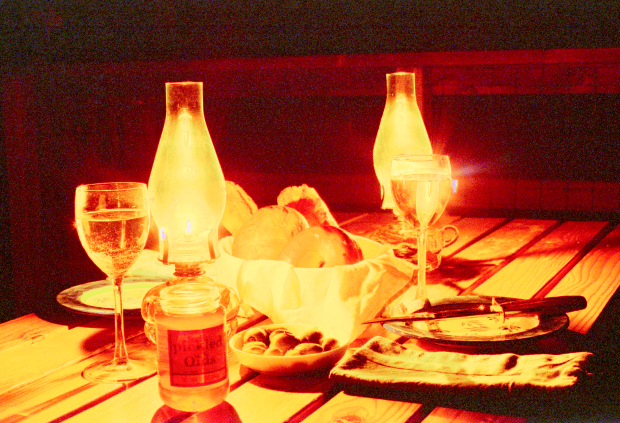}}%
\hspace{2mm}
	    \subfigure[Proposed
			\protect\newline(TMQI=0.8436, $\Delta \bm{c}$=0.0029, $\Delta \mathrm{H}$=2.5178)]{%
	        \includegraphics[clip, width=0.42\columnwidth]{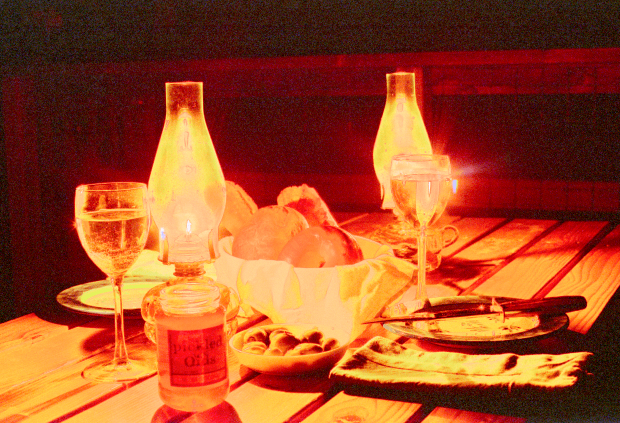}}%
\hspace{2mm}
	    \subfigure[Proposed with mean luminance adjustment
			\protect\newline(TMQI=0.8827, $\Delta \bm{c}$=0.0029, $\Delta \mathrm{H}$=2.9721)]{%
	        \includegraphics[clip, width=0.42\columnwidth]{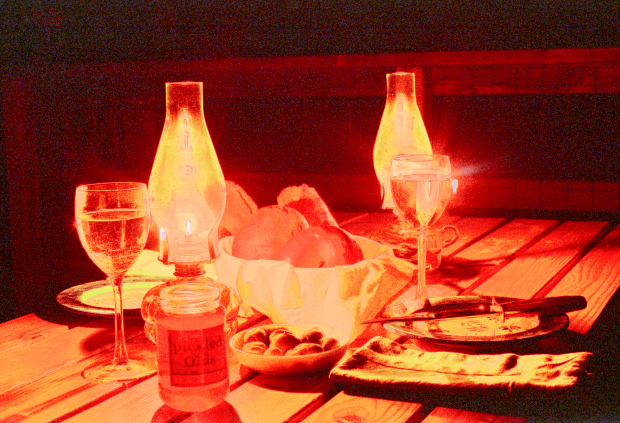}}%
	    \caption{Effects of mean luminance adjustment. (a)Tone-mapped image by Reinhard's global operator, (b)Tone-mapped image by using the proposed method, (c)Image adjusted from (b)}
\end{figure}

\bibliographystyle{ieicetr}
\bibliography{refs}

\begin{thebibliography}{10}

\bibitem{Reinhard}
E.~Reinhard, M.~Stark, P.~Shirley, and J.~Ferwerda, ``Photographic tone
  reproduction for digital images,'' ACM Trans. Graph. (TOG), vol.21, no.3,
  pp.267--276, 2002.

\bibitem{Drago}
F.~Drago, K.~Myszkowski, T.~Annen, and N.~Chiba, ``{Adaptive Logarithmic
  Mapping For Displaying High Contrast Scenes},'' Computer Graphics Forum,
  vol.22, no.3, pp.419--426, 2003.

\bibitem{Fattal}
R.~Fattal, D.~Lischinski, and M.~Werman, ``Gradient domain high dynamic range
  compression,'' ACM Trans. Graph., vol.21, no.3, pp.249--256, July\ 2002.

\bibitem{Shan}
Q.~{Shan}, J.~{Jia}, and M.S. {Brown}, ``Globally optimized linear windowed
  tone mapping,'' IEEE Transactions on Visualization and Computer Graphics,
  vol.16, no.4, pp.663--675, July\ 2010.

\bibitem{Gu}
B.~{Gu}, W.~{Li}, M.~{Zhu}, and M.~{Wang}, ``Local edge-preserving multiscale
  decomposition for high dynamic range image tone mapping,'' IEEE Transactions
  on Image Processing, vol.22, no.1, pp.70--79, Jan\ 2013.

\bibitem{Durand}
F.~Durand and J.~Dorsey, ``Fast bilateral filtering for the display of
  high-dynamic-range images,'' ACM Trans. Graph., vol.21, no.3, pp.257--266,
  July\ 2002.

\bibitem{Siku}
E.~{Sikudov^^c3^^a1}, T.~{Pouli}, A.~{Artusi}, A.O. {Aky^^c3^^bcz},
  F.~{Banterle}, Z.M. {Mazlumoglu}, and E.~{Reinhard}, ``A gamut-mapping
  framework for color-accurate reproduction of hdr images,'' IEEE Computer
  Graphics and Applications, vol.36, no.4, pp.78--90, July\ 2016.

\bibitem{tone1}
A.~Artusi, T.~Richter, T.~Ebrahimi, and R.K. Mantiuk, ``High dynamic range
  imaging technology [lecture notes],'' IEEE Signal Processing Magazine,
  vol.34, no.5, pp.165--172, Sep.\ 2017.

\bibitem{tone2}
A.~Artusi, F.~Banterle, T.~Ozan~Aydin, D.~Panozzo, and O.~Sorkine-Hornung,
  Image Content Retargeting: Maintaining Color, Tone, and Spatial Consistency,
  08\ 2016.

\bibitem{tone3}
T.~Murofushi, T.~Dobashi, M.~Iwahashi, and H.~Kiya, ``An integer tone mapping
  operation for hdr images for openexr with denormalized numbers,'' IEEE
  International Conference on Image Processing (ICIP), vol.97, no.11.

\bibitem{tone4}
T.~Dobashi, T.~Murofushi, M.~Iwahashi, and H.~Kiya, ``A fixed-point global tone
  mapping operation for hdr images in the rgbe format,'' IEICE Transactions on
  Fundamentals of Electronics, Communications and Computer Sciences, vol.97,
  no.11, pp.2147--2153, 2014.

\bibitem{tone5}
T.~Murofushi, I.~Masahiro, and H.~Kiya, ``An integer tone mapping operation for
  hdr images expressed in floating point data,'' IEICE Transactions on
  Fundamentals of Electronics, Communications and Computer Sciences, vol.97,
  no.11.

\bibitem{tone6}
Y.~Kinoshita, S.~Shiota, M.~Iwahashi, and H.~Kiya, ``An remapping operation
  without tone mapping parameters for hdr images,'' IEICE Transactions on
  Fundamentals of Electronics, Communications and Computer Sciences, vol.99,
  no.11.

\bibitem{Mantiuk}
R.~Mantiuk, R.~Mantiuk, A.~Tomaszewska, and W.~Heidrich, ``Color correction for
  tone mapping,'' Computer Graphics Forum, vol.28, no.2.

\bibitem{Yamaguchi}
Y.~Ueda, H.~Misawa, T.~Koga, N.~Suetake, and E.~Uchino, ``Hue-preserving color
  contrast enhancement method without gamut problem by using histogram
  specification,'' IEEE International Conference on Image Processing (ICIP).

\bibitem{Artit}
V.~Artit, Y.~Kinoshita, and H.~Kiya, ``Pure-color preserving multi-exposure
  image fusion,'' Proc. International Workshop on Advanced Image Technology,
  pp.657--667, Jan\ 2019.

\bibitem{ie1}
C.L. Chien and D.C. Tseng, ``Color image enhancement with exact hsi color
  model,'' International Journal of Innovative Computing, Information and
  Control, vol.7, pp.6691--6710, 12\ 2011.

\bibitem{ie2}
S.K. Naik and C.A. Murthy, ``Hue-preserving color image enhancement without
  gamut problem,'' IEEE Transactions on Image Processing, vol.12, no.12,
  pp.1591--1598, Dec\ 2003.

\bibitem{ie3}
M.~Nikolova and G.~Steidl, ``Fast hue and range preserving histogram
  specification: Theory and new algorithms for color image enhancement,'' IEEE
  Transactions on Image Processing, vol.23, no.9, pp.4087--4100, 2014.

\bibitem{ie4}
K.~Murahira and A.~Taguchi, ``Hue-preserving color image enhancement in rgb
  color space with rich saturation,'' pp.266--269, Nov\ 2012.

\bibitem{ie5}
S.~Yang and B.~Lee, ``Hue-preserving gamut mapping with high saturation,''
  Electronics Letters, vol.49, no.19, pp.1221--1222, Sep.\ 2013.

\bibitem{ie6}
Y.~Yu, Y.~Zhang, and S.~Yuan, ``Quaternion-based weighted nuclear norm
  minimization for color image denoising,'' Neurocomputing, vol.332, pp.283 --
  297, 2019.

\bibitem{CIEDE2000}
M.R. Luo, G.~Cui, and B.~Rigg, ``The development of the cie 2000
  colour-difference formula: Ciede2000,'' Color Research \& Application,
  vol.26, no.5, pp.340--350, 2001.

\bibitem{TMQI}
Y.~Hojatollah and W.~Zhou, ``Objective quality assessment of tone-mapped
  images,'' IEEE Transactions on Image Processing, vol.22, no.2, pp.657--667,
  Feb\ 2013.

\bibitem{HDR}
``{High dynamic range image examples}.'' http:/\slash{}www.\linebreak
  anyhere.com\slash{}gward\slash{}hdrenc\slash{}pages\slash{}originals.html.

\bibitem{SSIM}
Z.~Wang, A.C. Bovik, H.R. Sheikh, and E.P. Simoncelli, ``Image quality
  assessment: from error visibility to structural similarity,'' IEEE
  Transactions on Image Processing, vol.13, no.4, pp.600--612, April\ 2004.

\end{thebibliography}



\newpage

\profile[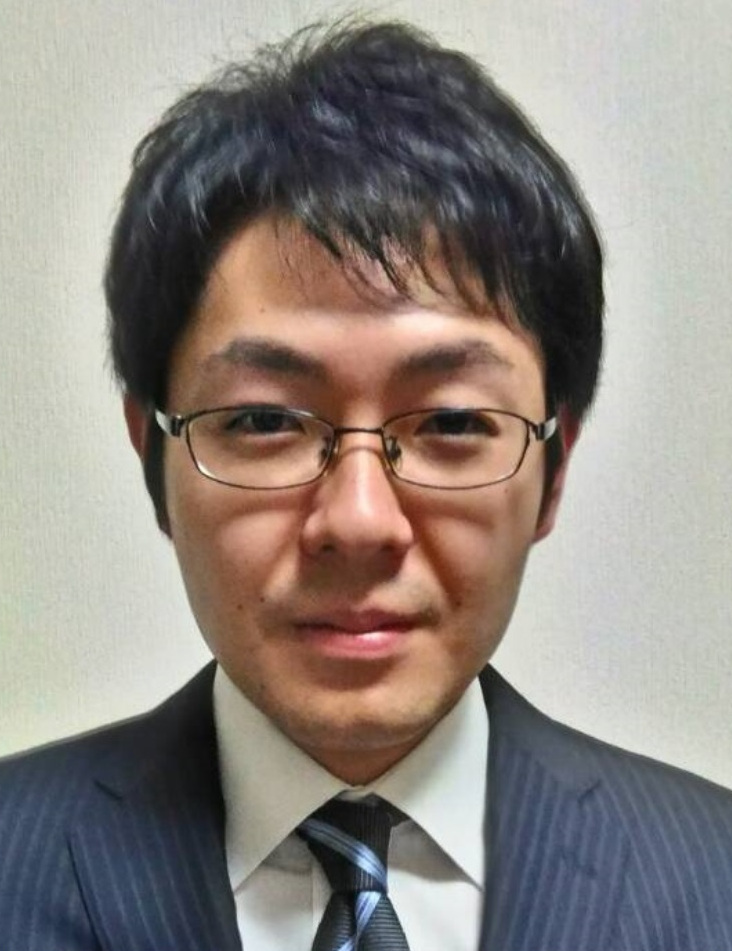]{Yuma Kinoshita}{received his B.Eng. and M.Eng. degrees from Tokyo Metropolitan University, Japan, in 2016 and 2018, respectively. From 2018, he has been a Ph.D. student at Tokyo Metropolitan University. He received IEEE ISPACS Best Paper Award in 2016, IEEE Signal Processing Society Japan Student Conference Paper Award in 2018, and IEEE Signal Processing Society Tokyo Joint Chapter Student Award in 2018, respectively. His research interests are in the area of image processing. He is a student member of IEEE and IEICE.}

\profile[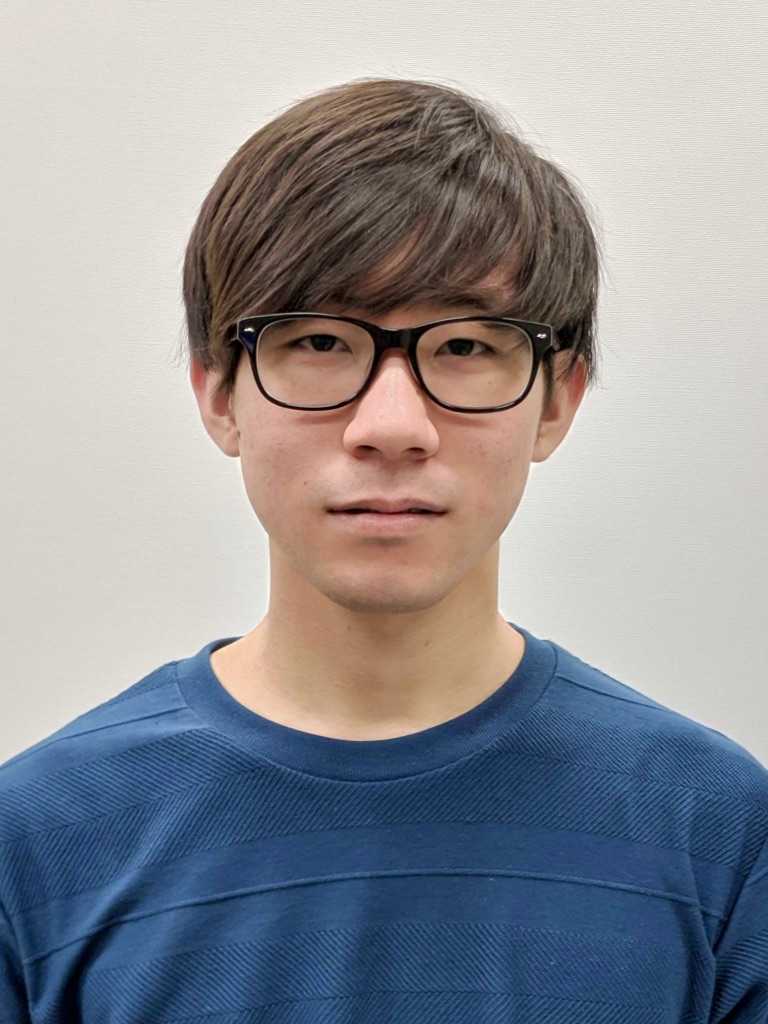]{Kouki Seo}{has been a Beginer course student at Tokyo Metropolitan University from 2015. His research interests include image processing.}

\profile[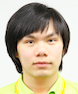]{Artit Visavakitcharoen}{received his B.Eng. and M.Eng. degree from King Mongkut's University of Technology Thonburi, Thailand in 2013 and 2015, respectively. He currently studies in Ph.D. course at Tokyo Metropolitan University, Japan. His research interests include image processing.}

\profile[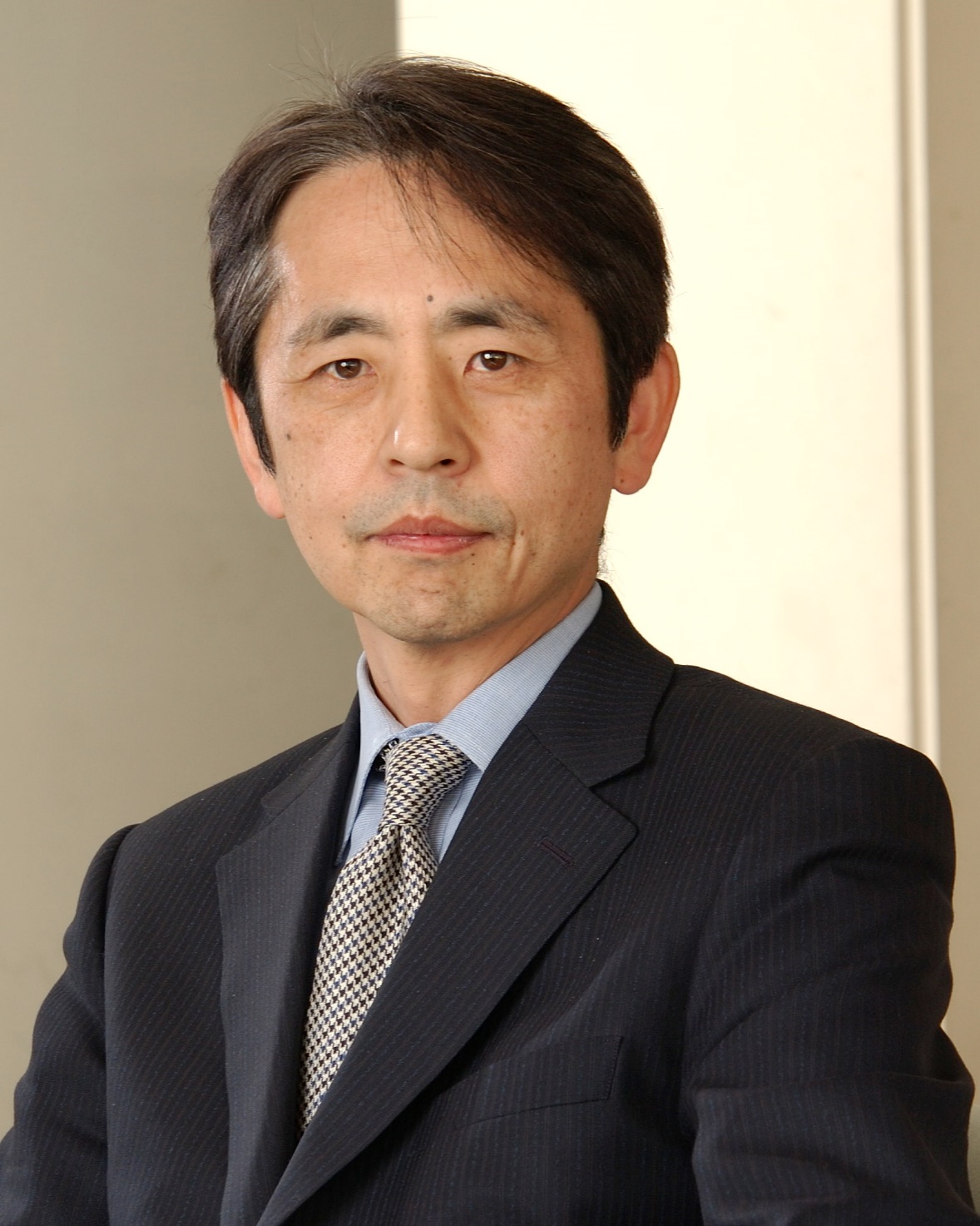]{Hitoshi Kiya}{received his B.E and M.E. degrees from Nagaoka University of Technology, in 1980 and 1982 respectively, and his Dr. Eng. degree from Tokyo Metropolitan University in 1987. In 1982, he joined Tokyo Metropolitan University, where he became Full Professor in 2000. From 1995 to 1996, he attended the University of Sydney, Australia as a Visiting Fellow. He is a Fellow of IEEE, IEICE and ITE. He currently serves as President of APSIPA, and he served as Inaugural Vice President (Technical Activities) of APSIPA in 2009-2013, and as
Regional Director-at-Large for Region 10 of IEEE Signal Processing Society in 2016-2017. He was also President of IEICE Engineering Sciences Society in 2011-2012, and he served there as Vice President and Editor-in-Chief for IEICE Society Magazine and Society Publications. He was Editorial Board Member of eight journals, including IEEE Trans. on Signal Processing, Image Processing, and Information Forensics and Security, Chair of two technical committees and Member of nine technical committees including APSIPA Image, Video, and Multimedia Technical Committee (TC), and IEEE Information Forensics and Security TC. He has organized a lot of international conferences, in such roles as TPC Chair of IEEE ICASSP 2012 and as General Co-Chair of IEEE ISCAS 2019. Dr. Kiya is a recipient of numerous awards, including six best paper awards.}

\end{document}